# A quantum spectrometer using a pair of phase-controlled spatial light modulators for superresolution in quantum sensing

Byoung S. Ham[1,2]

[1]School of Electrical Engineering and Computer Science, Gwangju Institute of Science and Technology, 123 Chumdangwagi-ro, Buk-gu, Gwangju 61005, South Korea
[2]Qu-Lidar, 123 Chumdangwagi-ro, Buk-gu, Gwangju 61005, South Korea
(June 03, 2024; bham@gist.ac.kr)

**Abstract**
Superresolution is a unique quantum feature generated by N00N states or phase-controlled coherent photons via projection measurements in a Mach-Zehnder interferometer (MZI). Superresolution has no direct relation with supersensitivity in quantum sensing and has a potential application for the precision measurement of an unknown signal's frequency. Recently, phase-controlled quantum erasers have been demonstrated for superresolution using classical light of a continuous-wave laser to overcome the diffraction limit in classical physics and to solve the limited scalability in N00N state-based quantum sensing. Here, a quantum spectrometer is presented for the macroscopic superresolution using phase-controlled spatial light modulators (SLMs) in MZI. For validity, a general solution of the superresolution is derived from the SLM-based projection measurements and an unprecedented resolution is numerically confirmed for an unknown frequency of light.

## 1 Introduction

A high-precision measurement has been an important research topic in science and engineering [1-10]. The minimum measurement error in classical physics is known as the standard quantum limit or the shot-noise limit (SNL) determined by the uncertainty principle of quantum mechanics [11]. In an interferometer, the phase sensitivity defines the ultimate phase error for an unknown signal [12]. In classical physics, the Rayleigh criterion defines the diffraction-limited resolution of an optical signal [13]. In the coherence optics of an N-slit (or grating) interferometer, the resolution increases as $\pi/N$, where the interference fringes are fixed at $\pm m\pi$ in the phase space regardless of N [13]. Unlike superresolution in quantum sensing [14-18], the overall phase sensitivity of the N-slit interferometer has no N gain due to the N-independent fixed fringes. Quantum sensing and metrology have been studied for both supersensitivity and supersensitivity to overcome the classical limit of SNL [14-22], where nonclassical lights of entangled photons [16-18] and squeezed states [20-22] are essential elements. N00N states have been intensively studied for superresolution in a Mach-Zehnder interferometer (MZI) [23-26], where squeezed states have no benefit. Although the superresolution has N gain in both resolution and phase sensitivity [15], the achieved N number in N00N states is far less than 100, so the implementation of the quantum sensor for superresolution has been severely halted.

Coherent single photons have also been studied to demonstrate superresolution in a noninterferometric system [27,28]. Recently, macroscopic superresolution has been proposed [29] and experimentally demonstrated [30] in an MZI using a continuous wave (cw) laser. For this, the concept of coherent single-photon-based quantum erasers [31] has been adapted for superresolution, where the quantum erasers are modified for equally spaced fringe shifts [29,30]. For this, a polarization-dependent phase delay of a quarter-wave plate (QWP) is used for the polarization-basis projection measurement [29]. Finally, a universal scheme of phase-controlled quantum erasers has been applied for scalable superresolution in a macroscopic regime [32]. For this, one MZI output port is divided into phase-controlled N ports, resulting in the equally spaced fringe shift. Thus, the intensity product of all phase-controlled quantum erasers results in superresolution [32]. Here, a quantum spectrometer is proposed for a potential application of generalized superresolution [32] to overcome the classical limit and to realize quantum sensing. For this, a pair of spatial light modulators (SLMs) are manipulated for the phase-controlled quantum erasers to replace the bulky optics of beam splitters (BSs) and QWPs in the original scheme [32]. Because MZI has already been well developed for Si-photonics as a key element [33], a chip-scale quantum spectrometer could be also possible [34].

## 2 Results

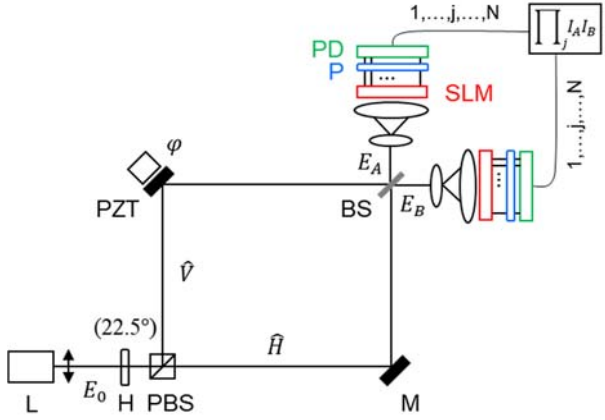

**Fig. 1.** Schematic of a quantum spectrometer using phase-controlled SLMs for superresolution. L: Laser, H: Half-wave plate, PBS: Polarizing beam splitter, M: Mirror, PZT: piezo-electric transducer, BS: nonpolarizing 50/50 beam splitter, PD: 1D or 2D photodiode, SLM: 1D or 2D spatial light modulator, P: 45° polarizer, $I_j = E_j E_j^*$.

Figure 1 shows the schematic of the proposed quantum spectrometer (or quantum wavelength meter) based on the macroscopic superresolution theory [32] using a synchronized SLM pair. The pixel-to-pixel arrayed SLM pair in Fig. 1 is either one- or two-dimensional for joint cross-correlations. For the polarization-sensitive phase delay, each SLM pixel pair is individually but jointly voltage-controlled to induce the phase shift given by $\xi_j = \pi j/N \ (j = 1, 2, \dots N)$.

The role of SLMs is to conduct quantum erasers together with polarizers via polarization-basis projection measurements [35]. Like QWP, SLM is a polarization-sensitive birefringent material. As demonstrated with QWP [30], the slow-axis vertical (or fast-axis horizontal) of SLM is the reference to result in a phase delay to the vertically polarized light, resulting in the fringe shift in the corresponding quantum eraser [29,30]. Using the BS matrix in coherence optics [36], the amplitudes of the MZI output fields in Fig. 1 are represented by:

$$\boldsymbol{E_A} = \frac{E_0}{\sqrt{2}}\left(\hat{H} - \hat{V}e^{i\varphi}\right), \tag{1}$$

$$\boldsymbol{E_B} = \frac{iE_0}{\sqrt{2}}\left(\hat{H} + \hat{V}e^{i\varphi}\right), \tag{2}$$

where $E_0$ is the amplitude of the input light L. $\hat{H}$ and $\hat{V}$ are the unit vectors of the horizontal and vertical polarization bases of $E_0$, respectively. To randomly generate the orthogonal polarization bases, the coherent light $E_0$ is passed through a 22.5°-rotated half-wave plate H before entering MZI. Due to the orthogonal polarization bases deterministically assigned to MZI paths, Eqs. (1) and (2) cannot contribute to interference fringes by the Fresnel-Arago law in coherence optics [37]. In a single-photon regime of quantum mechanics, this noninterfering phenomenon is known as distinguishable photon characteristics of a single photon in a viewpoint of the particle nature, though [38]. As studied for the quantum eraser [39], projection measurements of the polarization-controlled photons onto the polarizer P can retrospectively change the photon's nature from the particle (distinguishable) to the wave (indistinguishable), resulting in interference fringes [31]. Here, the quantum eraser has been coherently interpreted as a selective measurement processor at the cost of 50% measurement-event loss using the wave nature of a single photon [29-32].

In Fig. 1, the arrayed and synchronized SLM pair takes over the role of QWPs in the original scheme [32] to provide such a discrete phase relation among the SLM pixel-controlled individual lights. As a result, the same out-of-phase N quantum-eraser sets are generated via the polarization projection onto the common polarizer P. The out-of-phase relation in all QWP-controlled quantum eraser sets [30] is replaced by any jth pixel pair of the arrayed SLMs, resulting in Eqs. (3) and (4). For the beam expansion, a telescope composed of a lens pair is used to fit the 2D array of SLM, as shown in Fig. 1. For a 1D SLM, a cylindrical lens can be used, instead. To individually and independently detect the N-arrayed phase-controlled lights for polarization-basis projection through P, a corresponding arrayed pair of synchronized photodetectors (PDs) is followed. Thus, the jth pair of phase-controlled lights detected by the PD pair is represented as follows via polarization-basis projection onto the polarizers rotated at $\theta = 45°$ from Eqs. (1) and (2) (See refs. 29, 30, and 32 for the phase-controlled quantum eraser.):

$$\langle I_A^j \rangle = \frac{I_0}{2N}\langle 1 - \cos\psi_j \rangle, \tag{3}$$

$$\langle I_B^j \rangle = \frac{I_0}{2N}\langle 1 + \cos\psi_j \rangle, \tag{4}$$

where $\psi_j = \varphi - \xi_j$. $\xi_j = \pi j/N$ is the discrete phase control of the jth pixel pair of SLMs. The arrayed phase control of SLM does not have to be dynamic due to the fixed phase relation $\xi_j$ [32].

Regarding recently demonstrated phase-controlled superresolution using a commercial laser [30], the linear optics-based phase-control system is too bulky to satisfy the scalability for N>100. Thus, the superresolution $\pi/N$ is potentially limited to beat the classical counterpart [40]. For the N scalability, a pair of SLMs replaces both BSs and QWPs for the arrayed quantum erasers [32]. As shown by the general solution of the phase-controlled superresolution [32], the discrete phase assignment to the SLM pixels results in the same fringe shifts between arrayed light fields by Eqs. (3) and (4). For this, the jth pixel pair of SLMs is phase-controlled for $\xi_j = \pi j/N$ (see Eq. 25 in ref. 32). All phase-controlled lights from the SLM pair are individually detected by the one-to-one corresponding photodiode arrays. As shown in Eqs. (3) and (4), the jth SLM-controlled output fields satisfy the out-of-phase relation automatically for the MZI output ports (see Eqs. (1) and (2)). Thus, the jth intensity product between jth arrays of the PD pair is represented by:

$$\langle R_{AB}^{(j)} \rangle = \frac{I_0^2}{2^2}\langle \sin^2(\varphi - \xi_j) \rangle. \tag{5}$$

As a result, the Nth-order intensity correlation between all SLM-controlled light arrays is as follows:

$$\langle R_{AB}^{2N} \rangle = \frac{I_0^{2N}}{2^{2N}}\langle \prod_{j=0}^N \sin^2(\varphi - \xi_j) \rangle, \tag{6}$$

For an arbitrary order N, Eq. (6) satisfies the same phase-controlled superresolution [32], representing a macroscopic quantum spectrometer (or quantum wavelength meter). Here, it should be noted that Eq. (6) is actually for 2N intensity correlations because of the $\sin^2(\varphi - \xi_j)$ between SLM pixel-controlled light pairs for the second order (see the order 2N in $R_{AB}^{2N}$).

To verify the SLM-induced general solution of the superresolution in Eq. (6), numerical calculations are conducted in Fig. 2. The magnitudes of $\langle R_{AB}^{2N} \rangle$ are normalized to compare the resolution $\Delta_{2N}$. By the N-proportional fringe numbers of superresolution as discussed in ref. [32], the resolution of the Nth-order intensity correlation satisfies the Heisenberg limit of quantum sensing, resulting in $\Delta_N = \pi/N$ according to the Rayleigh criterion [13]. In the top row of Fig. 2, two-pixel pairs of SLMs for j=0 and 1 are shown for N=2. This case is for a two-pixel SLM in each MZI output port. As shown in the top left and middle panels, each pixel pair (see Section A of the Supplementary Material) and their intensity product show the out-of-phase relation as in a typical MZI. Thus, the intensity product $R_{AB}^{2N}$ between them results in doubled fringe numbers, as shown in the top right panel. Compared with the Heisenberg limit $\pi/N$, it shows the same enhanced resolution, $\Delta_{2N} = \pi/2N$, where only 2N/2 photons are involved in each path of MZI.

In the middle row, SLM pixels increase to ten. Thus, the jth pixel is adjusted with $\xi_j = \pi j/10$, as indicated by arrows for j=2 and 7 in the middle-left panel. The dotted curves in the center panel show the corresponding intensity products (detected by the PD pair). Thus, the intensity product $R_{AB}^{20}$ between all ten paired SLM pixels results in a five-times increment in fringe numbers over the top row, as shown in the right-end panel of the middle row. As a result, the resolution is enhanced five times compared with the first row, $\Delta_{20} = \pi/20$, satisfying the Heisenberg limit.

The bottom row is for ten times more pixel numbers than the middle row. Thus, the jth pixel pair of SLMs is phase controlled by $\xi_j = \pi j/100$. The arrows in the bottom-left panel show the 20th and 70th pixels of the SLM pair. The dotted curves in the middle panel of the bottom row are for the corresponding pixel-resulting intensity products (detected by joint PDs). Compared with the middle row, the number of fringes increases ten times. For all 100 pixel pairs of SLMs, the intensity product $R_{AB}^{200}$ shows ten times increased fringe numbers compared with that in the middle row, resulting in the enhanced resolution to $\Delta_{200} = \pi/200$. Thus, the proposed intensity product method by the phase-controlled SLM pair confirms the enhanced resolution $\Delta_{2N} = \pi/(2N)$. This resolution enhancement satisfies the Heisenberg limit in the conventional N00N-based quantum sensing [11,12,14-18,23-26] and phase-controlled superresolution [27-30,32].

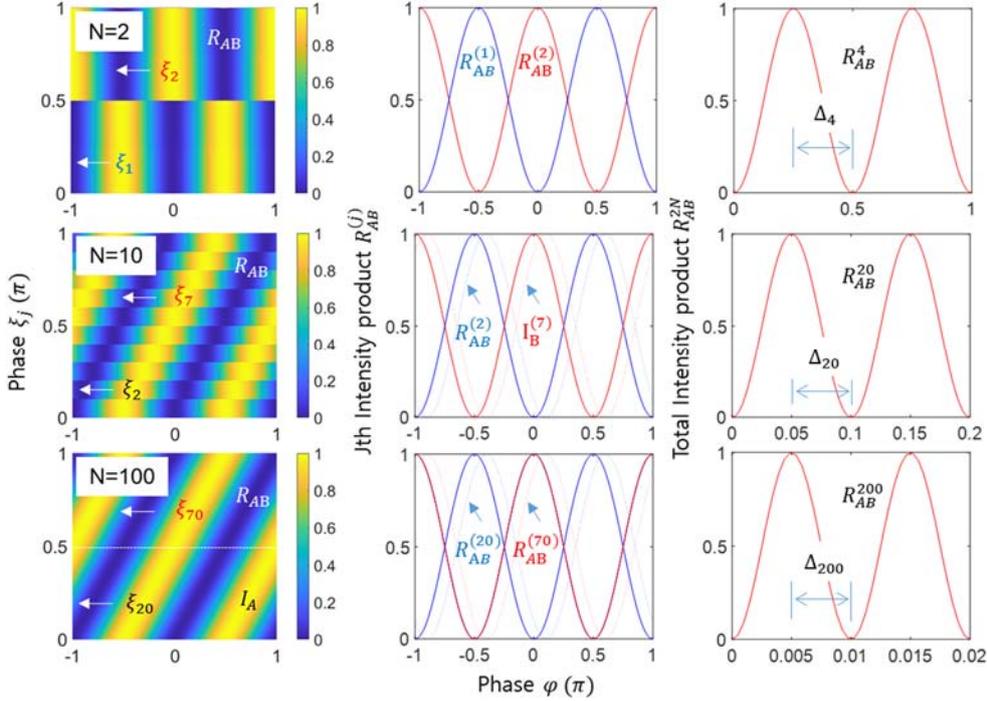

**Fig. 2.** Numerical calculations of the intensity product for Eqs. (3), (4), and (6).

In Fig. 3, phase sensitivity is analyzed for Fig. 2. According to the superresolution analyzed in Fig. 2 for Fig. 1, the phase sensitivity $\delta\varphi$ is dependent upon $\Delta$: $\delta\varphi = \delta I/(\frac{dI}{d\varphi})$ [11,32]. Compared with fixed fringes at $\pm m\pi$ (m=0,1,2...) in classical N-slit or N-groove grating interferometer [13], the superresolution of Eq. (6) gives a unique and distinctive quantum feature of $I^N \propto sin^2(N\varphi)$ over the classical counterpart [11], as numerically demonstrated in Fig. 2. Regarding the resolution at $\varphi = \pm m\pi$ (m=0,1,2...), the N-slit interferometer satisfies the same Heisenberg limit via N-wave superposition, $\Delta_{N-slit} = \pi/N$, though (see Section B of the Supplementary Materials): $I^1 = \mathrm{Sinc}^2\beta(sin(N\varphi)/sin\varphi)^2$ [13]. Thus, the phase sensitivity of the N-slit interferometer also has the same gain effect as in the superresolution but only at the fringe positions: The overall phase sensitivity of the N-slit interferometer has no effective gain for an unknown signal due to nulls in most phase variations. On the contrary, the superresolution of Eq. (6) has the N times increased effective frequency $f_{effc} = Nf_0$, resulting in N times increased fringe numbers. For details of calculations, Fig. 3 shows the slope $dR_{AB}^{2N}/d\varphi$ (blue curve) of Fig. 2 for a unit phase period $-\pi/2 \leq \varphi \leq \pi/2$ together with that of N-slit interferometer (red curves). Figure 3(d) is for the overall slopes averaged.

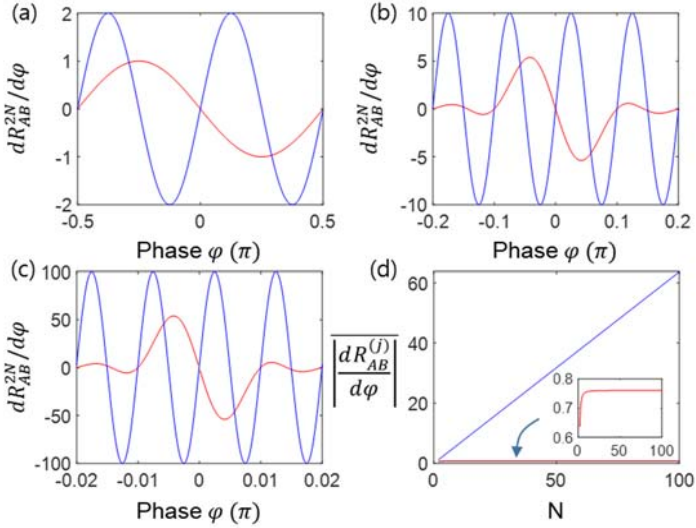

**Fig. 3.** Numerical calculations of derivatives. (a) N=2. (b) N=10. (c) N=100. (d) The mean slope for a unit phase period. (blue) $dR_{AB}^{2N}/d\varphi$. (red) $d(sin(N\varphi)/sin\varphi)^2/d\varphi$.

The red curves in Fig. 3 are for the effective phase sensitivity of the N-slit or N-groove interferometer applied for the conventional spectrometer or wavelength meter [13,40]. The traditional wavelength meter based on a Fabry-Perot interferometer shows the same feature due to the effective N by reflection-coefficient-based Finesse (see Section C of the Supplementary Materials) [13]. Due to the fixed fringes at $\varphi = \pm m\pi$, the slope of the phase sensitivity is proportional to N, but compensates for the line narrowing, resulting in the flat gain feature, as shown in the Inset of Fig. 3(d).

On the contrary, the derivative (blue curves) of Eq. (6) shows N repeated pattern as shown in Figs. 3(a)-(c), resulting in a linearly increased effective phase sensitivity (see the blue line in Fig. 3(d)). This is proof of the effective gain in phase sensitivity, as demonstrated with photonic de Broglie waves in quantum sensing [14-18,23-26]. Here it should be noted that superresolution (phase sensitivity) has nothing to do with supersensitivity (amplitude sensitivity), as shown, e.g., by squeezed light [21,22] and N00N states [15]. Although the physics between the classical and quantum interferometers is different in the order of intensity correlation, the same number of discrete fields contributes to the same resolution satisfying the Heisenberg limit (see Sections B and C of the Supplementary Materials). Thus, the discrete phase relation between paired SLM pixels is the bedrock of the quantum feature, as discussed for phase quantization [32,41]. On the contrary, the classical phase sensitivity of the N-slit interferometer or grating-based spectrometer has no quantum gain due to the fixed fringes at $\pm m\pi$ (see the Inset of Fig. 3(d)).

Figure 4 shows numerical simulations of frequency resolution for the proposed quantum spectrometer using SLM-based superresolution. Equation (6) is used for the analysis. In Fig. 4, the frequency difference $\delta f$ induces the phase $\delta\varphi = \delta f \tau$, where $\tau$ is the path-length difference-resulting temporal delay in MZI of Fig. 1. Thus, the photodetector measures a beating frequency $\Delta f$ between reference frequency $f_0$ (green curves) and an unknown frequency $f$. From the left to the right columns in Fig. 4, the pixel number of each SLM increases from 2 to 10, 100, and 10,000, respectively. A 1024x1272 pixel SLM is a common out-of-shelf product, so N=10,000 is acceptable. The beating frequency is easily captured by PDs for the intensity-product order N in Eq. (6).

For N=2 in the left-end column, $\delta f = \pm 0.1 f_0$ is set for an unknown $f$ (see the blue and red arrows). For the scan range $-10\pi \leq \varphi \leq 10\pi$, Eq. (6) results in 40 fringes, as shown by the green curve in the lower left-end panel. Due to the $\pm 0.1 f_0$, the number of beating fringes should be 10%, i.e., 4, as shown by the red and blue envelopes. For N=10 in the middle-left column, the beating frequency should increase by five times, resulting in 20 beating fringes, as shown by the red and blue envelopes: The scan range is decreased by ten times, resulting in two corresponding beating fringes. Likewise, for N=100 in the middle-right column, the beating fringes must be the same as that of the middle-left column to compensate $\delta f = \pm 0.01 f_0$. In the right-end column, the beating fringes must be 100 times increased compared to the middle-right column (see the 100 times decreased scan range). Thus, the frequency resolution by Eq. (6) is linearly proportional to the SLM pixel number N, satisfying the Heisenberg limit $\Delta = \pi/N$. This effect is critical for $N \gg 10^4$ in the proposed method compared with the conventional grating-based spectrometer limited by $N \sim 10^4/cm$. Moreover, the proposed phase-controlled SLM-pair-based quantum spectrometer works for MZI via an intensity product between all SLM pixels. Finding an unknown frequency from the measurement of the Nth-order intensity product is conducted by the frequency beating process. Considering $N_{max} = 10^4$ in the conventional spectrometer, the proposed quantum spectrometer can give a hundred times enhanced resolution and phase sensitivity because a $10^6$-pixel SLM is commercially available. In that sense, not only a quantum spectrometer but also a quantum gyroscope can be a potential candidate for the proposed method (discussed elsewhere).

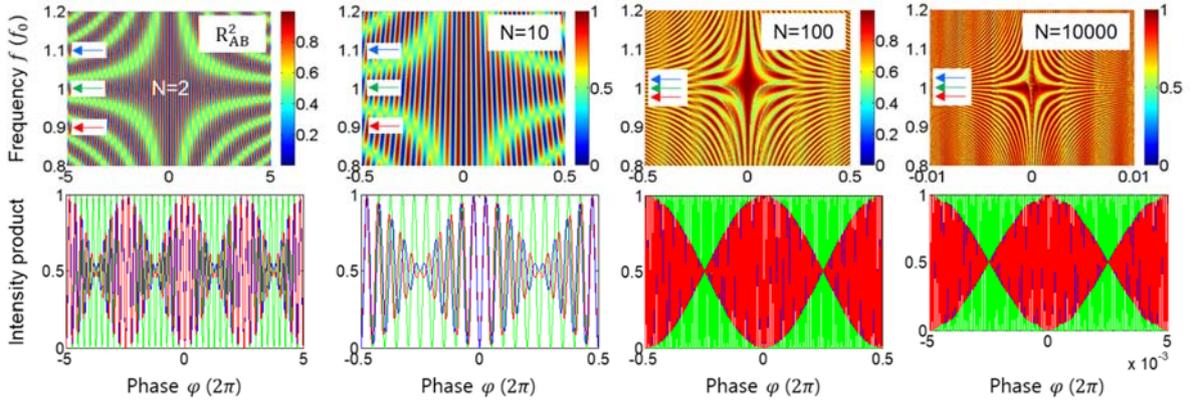

**Fig. 4.** Numerical calculations for the frequency resolution of an unknown signal. The colored arrows in the top row show beat frequencies in the bottom row. For N=2 and 10 , $f^{\pm} = (1 \pm 0.1) f_0$. For N=100 and 10,000, $f^{\pm} = (1 \pm 0.01) f_0$.

## 3 Conclusion

Recently, a proposed coherent-state-based superresolution theory [29,32] was applied for a potential quantum spectrometer (or wavelength meter) using a phase-controlled SLM pair. To replace the bulky BS-QWP structure in the original scheme [32], an arrayed SLM pair gave a practical solution of the N scalable quantum erasers satisfying the out-of-phase relation between each SLM pixel pair. For the intensity-correlation measurements between phase-controlled SLMs, an arrayed photodiode was aligned via a 45-degree-rotated polarizer in each MZI output. For the discrete phase relation between all pixels of SLMs, the applied voltages to SLMs were correspondingly discrete, where no dynamic control was needed. Unlike conventional grating (N-slit)-based spectrometer, the proposed SLM-based quantum spectrometer satisfied the superresolution, enhancing phase sensitivity N times for an unknown signal due to the N increased fringes. A general solution of the SLM-based quantum spectrometer was derived and numerically confirmed for superresolution and compared with a conventional grating (or N-slit)-based-spectrometer. As discussed for the phase quantization in the superresolution [32,41], the discrete phase relation between N coherent fields resulted in the same resolution as in the N-slit system. Unlike conventional spectrometers, however, the N-increased fringes in the proposed quantum spectrometer showed a quantum advantage in phase sensitivity for an unknown signal. Moreover, when an out-of-shelf SLM is used, the proposed quantum spectrometer potentially showed a hundred times-enhanced resolution for an unknown frequency.

## Methods

Each SLM pixel pair plays the role of a QWP-controlled quantum eraser set for the out-of-phase relation [29,32]. For the polarization-basis projection measurement of the SLM pixel pair, random polarization bases of the input light are provided by a 22.5-degree rotated half-wave plate H. Thus, the photon characteristics inside the MZI satisfies the particle nature by the polarizing beam splitter, resulting in no interference fringes in $I_A$ and $I_B$. In Fig. 1. For the intensity-product measurement between all SLM pixel pairs, each photodiode is pixel-to-pixel aligned to each SLM, as shown in Fig. 1. The polarizer P is sandwiched between SLM and PD. Finally, an electronic circuit works real-time intensity product calculations between all N-paired PD pixels.

## Disclosures
The author declares no competing interest.

## Data Availability
All data generated or analyzed during this study are included in this published article.

## Author contribution

BSH solely wrote the paper.

### Acknowledgments
This research was supported by the MSIT (Ministry of Science and ICT), Korea, under the
ITRC (Information Technology Research Center) support program (IITP 2024-2021-0-01810) supervised
by the IITP (Institute for Information & Communications Technology Planning & Evaluation). This work was
also supported by GIST via the GIST research program in 2024.